# New Physics Data Libraries for Monte Carlo Transport

M. Augelli, S. Hauf, M. Kuster, M. Han, C. H. Kim, M. G. Pia, L. Quintieri, H. Seo, P. Saracco, G. Weidenspointner and A. Zoglauer

*Abstract*–The role of data libraries as a collaborative tool across Monte Carlo codes is discussed. Some new contributions in this domain are presented; they concern a data library of proton and alpha ionization cross sections, the development in progress of a data library of electron ionization cross sections and proposed improvements to the EADL (Evaluated Atomic Data Library), the latter resulting from an extensive data validation process.

## I. Introduction

Data libraries, consisting of tabulations of physics quantities originating from experimental or theoretical sources, are a widely used, essential tool in Monte Carlo simulation.

Their main purpose is to facilitate the simulation of physics processes by providing evaluated compilations of experimental data (or fits to them), or tabulations of theoretical quantities, which would be cumbersome to perform in the course of the simulation execution.

Data libraries could also play another valuable role in the context of Monte Carlo simulation application and benchmarking. Tabulations of fundamental physics quantities (cross sections, secondary particle spectra etc.) used by a Monte Carlo code could be a means for evaluating the effects of different physics modeling approaches in the same simulation environment of particle transport, geometry, user-defined cuts etc. The public provision of such tabulations should be promoted in the scientific community to facilitate the evaluation of possible systematic effects in simulation application results and to contribute to simulation validation efforts.

Recent activity concerning the creation of new data libraries meant for public distribution is reviewed in the following sections. Recent physics validation results, which question the accuracy of currently available data tabulations, are also discussed and suggestions for their update are proposed to better reflect the state-of-the-art in the associated field.

Manuscript received November 17, 2010.
M. Augelli is with CNES, Toulouse, France (e-mail: mauroaugelli@mac.com).
M. G. Pia and P. Saracco are with INFN Sezione di Genova, Via Dodecaneso 33, I-16146 Genova, Italy (telephone: +39 010 3536328, e-mail: MariaGrazia.Pia@ge.infn.it).
L. Quintieri is with INFN Laboratori Nazionali di Frascati, Frascati, Italy (e-mail: Lina.Quintieri@lnf.infn.it).
M. Han, H. Seo and C. H. Kim are with the Department of Nuclear Engineering, Hanyang University, Seoul 133-791, Korea (e-mail: mchan@hanyang.ac.kr; shee@hanyang.ac.kr; chkim@hanyang.ac.kr).
S. Hauf and M. Kuster are with TU Darmstadt, D-64289 Darmstadt, Germany (e-mail: steffen.hauf@skmail.ikp.physik.tu-darmstadt.de, markus.kuster@xfel.eu).
Georg Weidenspointner is with MPE, Garching, Germany, and with MPI Halbleiterlabor, Munich, Germany (e-mail: Georg.Weidenspointner@hll.mpg.de).
A. Zoglauer is with the Space Sciences Laboratory, University of California at Berkeley, Berkeley, CA 94720, USA (e-mail:zog@ssl.berkeley.edu)

## II. Proton and α Ionization Data Library

Recent progress in PIXE simulation [1] with Geant4 [2][3] involved the development of simulation tools for the calculation of ionization cross sections by proton and α particle impact based on a variety of theoretical and empirical approaches: the ECPSSR [4] model and its variants (with Hartree-Slater corrections [5], with the "united atom" approximation [6] and specialized for high energies [7]), theoretical plane wave Born approximation, empirical models based on fits to experimental data collected by Paul and Sacher [8], Paul and Bolik [9], Kahoul et al. [10], Miyagawa et al. [11], Orlic et al. [12] and Sow et al. [13]. The empirical models concern K shell ionization by protons and α particles, and L shell ionization by protons; the theoretical models concern K, L and M shell ionization by protons and α particles.

The tabulated cross sections have been subject to extensive validation with respect to compilations of experimental measurements [8][14][15]. The comparison process involved rigorous statistical methods [16][17] to estimate the compatibility between the tabulated cross sections and experiment, and to evaluate the relative accuracy among the various modeling options. The full set of validation results is documented in a dedicated paper [1].

The above mentioned cross sections have been tabulated in the energy range between 10 keV and 10 GeV; the tabulations of cross sections deriving from empirical models are limited to the energy range covered by the models themselves. The tabulations have been assembled in a data library, which is complemented by an example of basic software for retrieving the data and printing them. An example of the content of this data library is shown in Fig. 1.

The data library is intended for public distribution by RSICC (the Radiation Safety Information Computational Center at Oak Ridge National Accelerator Laboratory). The material for the data library (data files, documentation and example code to read them) has been submitted to RSICC shortly before the IEEE Nuclear Science Symposium 2010.

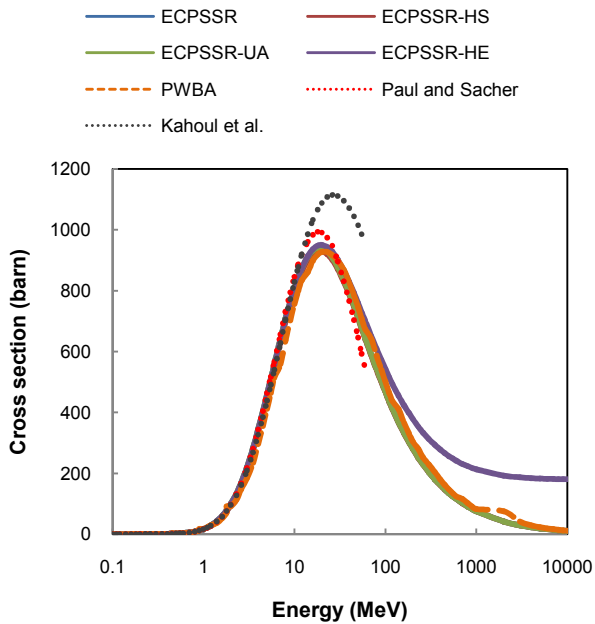

Fig. 1. Cross sections for copper K shell ionization by proton impact included in the data library.

III. ELECTRON IONIZATION DATA LIBRARY

Recent progress in the simulation of low energy electron ionization [18] with Geant4 involved the development of simulation tools for the calculation of ionization cross sections by electron impact based on the Binary-Encounter-Bethe (BEB) model [20] and the Deutsch-Märk (DM) model [21].

The implemented cross sections, along with the cross sections tabulated in the Evaluated Electron Data Library [22] (EEDL), have been subject to extensive validation with respect to a large collection of experimental measurements, including more than 100 individual data sets and concerning more than 50 target elements. The comparison process involved statistical methods [16][17] to estimate the compatibility between the calculated cross sections and experiment, and to evaluate the relative accuracy of the three modeling options. The full set of validation results is documented in a dedicated paper; a sample of results is presented in [18] [19].

The above mentioned cross sections have been tabulated in the energy range between 1 eV and 100 keV for elements with atomic number between 1 and 92 inclusive. The tabulations have been assembled in a data library, which is complemented by an example of basic software for retrieving the data and printing them. An example of the content of the data library is shown in Fig. 2, along with cross sections contained in the EEDL data library and deriving from calculations [23] also used in Penelope.

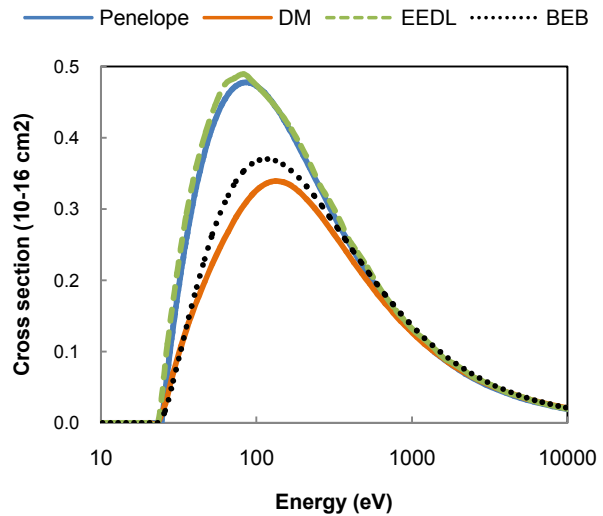

Fig. 2. Electron impact ionization cross sections for Helium as a function of energy; the BEB and DM cross sections contained in the data library are shown together with the cross sections tabulated in the EEDL library and deriving from calculations used in the Penelope code.

The data library is intended for public distribution; it will be released following the publication of the paper describing the new software developments and their validation. Since the validation process identified the new models as more accurate than EEDL at reproducing experimental measurements in the energy range below 250 eV, this data library could be a valuable alternative to EEDL in the lower energy range.

IV. IMPROVEMENT OF EADL

Recent analyses [24][25] for the experimental validation of parts of the Evaluated Atomic Data Library (EADL) [26] showed that some of its content does not reflect the state-of-the-art.

Regarding radiative transition probabilities, Hartree-Fock [27][28] calculations appear more accurate [24] than the Hartree-Slater [29][30] ones tabulated in EADL. Moreover, anomalies hinting to some accidental errors in the assembly of the library have been detected [31]; the flawed values are affected by errors amounting to orders of magnitude differences with respect to the original theoretical references [29][30] from which they derive.

The inner shell binding energies of and ionization energies tabulated in EADL have also been subject to validation with respect to experimental data. A set of results from this analysis concerning the accuracy of inner shell binding energies is presented in another paper [25]. A sample of results concerning ionization energies is shown in Fig. 3. The values in EADL appear in general less accurate than other compilations of electron binding energies available in the literature.

The full set of results deriving from this validation process is intended to be documented in a dedicated paper, whose publication will follow this conference.

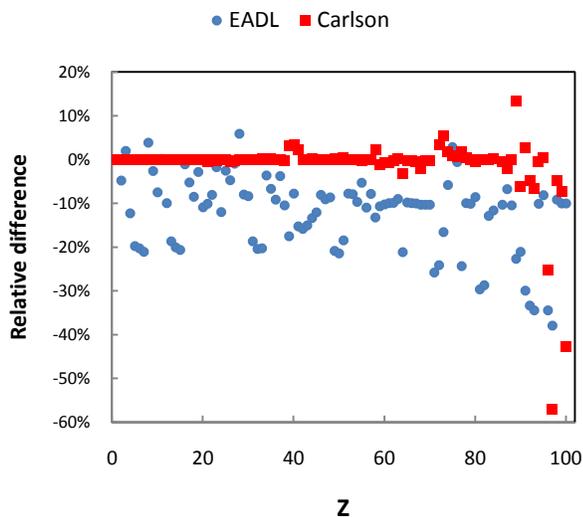

Fig. 3. Relative difference of ionization energies derived from EADL and Carlson's compilation with respect to NIST reference data (experimental).

These results suggest that an update of EADL would be beneficial to better reflect the state-of-the-art, which has evolved since the time of its last distributed version.

Along with the improvement of the content of EADL, a revision of its format to better match modern computational techniques would facilitate the use of this data library.

It should be stressed that EADL is an invaluable tool for Monte Carlo simulation, thanks to its wide collection of atomic parameters in one consistent environment; therefore it would be worthwhile to invest in its update.

## V. REVIEW OF RADIOACTIVE DECAY DATA

The quantitatively accurate representation of radioactive decays within a Monte Carlo simulation is of importance for a variety of applications such as dose calculations for medical, experimental and space flight purposes as well as estimating instrument responses in a wide range of experimental scenarios.

The current Geant4 radioactive decay simulation uses datasets which are based on the Evaluated Nuclear Structure Data Files (ENSDF) [32] to obtain half lives, decay branches, energy levels and level intensities of the decaying nucleus. It then passes the decayed nucleus to the photo-deexcitation process, which uses its own datasets to decay the nucleus into its ground state. The current database does not include references to the origins of the individual datasets or their actuality.

A comparison between the Geant4 datasets and the current version of the ENSDF shows disagreements in level energies for a considerable amount of isotopes as shown in Fig. 4.

There is currently no simple possibility for the end user to check the accuracy of the Geant4 data other than manually compare the entries to the state-of-the-art databases. An updated database which also includes references to the data origins would greatly facilitate this comparison. One should also consider providing decay data and deexcitation data as one consistent database, which would aid the identification of simulation inaccuracies in the common use case of a gamma ray detector detecting the deexcitation photons and not the levels of the decayed nucleus.

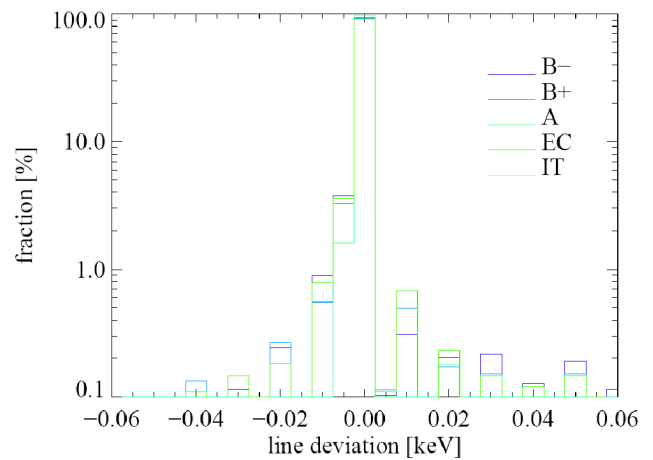

Fig. 4. A comparison between the Geant4 datasets and the current version of the ENSDF.

Finally an updated database could also make use of modern computational techniques such as an XML basis, which would facilitate parsing and structuring of the individual datasets.

## VI. CONCLUSION

Data libraries play an important role in Monte Carlo simulation.

New data libraries are currently in preparation, concerning ionization cross sections by electron, proton and $\alpha$ particle impact. Their accuracy has been estimated by means of extensive comparisons with experimental data.

Recent tests have shown the need of updating some parts of EADL to achieve better accuracy. Sources for such an improvement have been identified.

Provision of data libraries for open circulation within the scientific community should be promoted. Publicly available data libraries could facilitate comparisons of simulations based on different codes, as well as sharing physics modeling features in a variety of simulation environment.

The complete set of results are documented and discussed in depth in dedicated papers.

## ACKNOWLEDGMENT

The authors express their gratitude to CERN for support to the research described in this paper.

The authors thank Sergio Bertolucci, Elisabetta Gargioni, Simone Giani, Vladimir Grichine, Berndt Grosswendt, Andreas Pfeiffer and Alessandro Zucchiatti for valuable discussions, and Tullio Basaglia of the CERN Library for essential support to this study.

RSICC helpful collaboration for the assembly and distribution of the PIXE data library described in this paper is acknowledged.